\documentclass[a4paper]{jpconf}
\usepackage{graphicx}
\usepackage{hyperref}
\usepackage{enumitem}
\usepackage[scaled=0.85]{beramono}
\begin{document}
\title{CODE-RADE - Community Infrastructure for the Delivery of Physics Applications}

\author{Bruce Becker}
\address{C.S.I.R. Building 43 \\ 1 Meiring Naude Road \\ Brummeria \\ Pretoria \\ 0001}
\ead{bbecker@csir.co.za}

\author{Sean Murray}
\address{C.H.P.C. \\ Rosebank \\ Cape Town \\ 7700}
\ead{murrays@cern.ch}

\begin{abstract}
	Since the very first translation of a mathematical expression into binary executable code, the process of porting applications to computational resources has been at the core of scientific computing. Of the thousands of scholarly communications published every year, at the heart of the majority of them is the analysis of some data, the simulation of a physical, chemical, biological or other process and the interpretation thereof. Scientific computing can therefore in some sense be distilled to the execution of an application - or rather sets of applications which are combined into complex workflows. Due to the complexity and number both of scientific packages as well as computing platforms, delivering these applications to end users has always been a significant challenge through the grid era, and remains so in the cloud era. In this contribution we describe a platform for user-driven, continuous integration and delivery of research applications in a distributed environment - project CODE-RADE. Starting with 6 hypotheses describing the problem at hand, we put forward technical and social solutions to these. Combining widely-used and thoroughly-tested tools, we show how it is possible to manage the dependencies and configurations of a wide range of scientific applications, in an almost fully-automated way. The CODE-RADE platform is a means for developing trust between public computing and data infrastructures on the one hand and various developer and scientific communities on the other hand. Predefined integration tests are specified for any new application, allowing the system to be user-driven. This greatly accelerates time-to-production for scientific applications, while reducing the workload for administrators of HPC, grid and cloud installations. Finally, we will give some insight into how this platform could be extended to address issues of reproducibility and collaboration in scientific research in Africa.
\end{abstract}

	\section{Introduction}
	Since the first computing program exceeded the capabilities of its original execution host, and was attempted to be migrated to another platform whether it be bigger or in another building, scientists have suffered trying to get their research done. Countless hours have been wasted getting programs to simply run as opposed to actually doing the science they intended to do.

	It has been a long standing problem and irritation when one tries to get a program to run on computing infrastructure that is not sitting on your desk. The scale and complexity loosely follows that of the code and the computing system you wish to run on.

	\subsection{The paradox of  plentiful computing}

	It is a central theme of scientific computing that there are never enough resources available to any particular researcher picked at random from a sample. Simultaneously, however, there has been massive and near-pervasive investment in HPC and distributed computing resources. Indeed, there has been a coordinated national effort to build a well-connected distributed computing platform in South Africa\cite{SAGrid}, fully integrated and connected to the national research network\cite{SANREN}. This effort has even been extended at a trans-continental scale, via the signature of a memorandum of understanding between African and European infrastructure providers\cite{AAROC}. With the creation of the Africa-Arabia Regional Operations Centre\footnote{See \url{"http://www.africa-grid.org"} for details}, several sites across Africa are available to researchers to execute their scientific workflows, and support several different virtual organisations, or communities of practice. However, despite this effort in interoperability and integration, many of these resources are under-utilised at any given time. This seems to point to a paradox of scientific computing, where there is an over-provision of resources, yet scientists are still impeded from executing their calculations.

	\subsection{Barriers to Entry}

	It does not take much analysis of the situation to conclude that this is only seemingly a paradox, and the situation exists due to various barriers to entry, one of which is the ability to actually execute the code. Indeed one of the main benefits of a distributed computing platform is also one of the main barriers to entry : the complexity of the environment. Researchers seldom need to take into account the configuration of their application or workflow for different platforms, and it is practically unfeasible to insist on homogeneous hardware and configurations across the sites. This results in significantly increased overhead on the part of the researcher to ensure that their application will properly execute at any of the given resources, which thus leads to oversubscription at certain sites and underutilisation at others\footnote{Another effect is the reduction in the capacity of the researcher and experts at sites to effectively collaborate with each other, due to geographical and technical divergence.}.

	CODE-RADE aims to reduce or remove the barrier to entry for new applications, thus eliminating the paradox of plentiful computing.

	\section{CODE-RADE Design and Philosophy}

\begin{figure}[h]
		\begin{center}
			\includegraphics[width=0.75\textwidth]{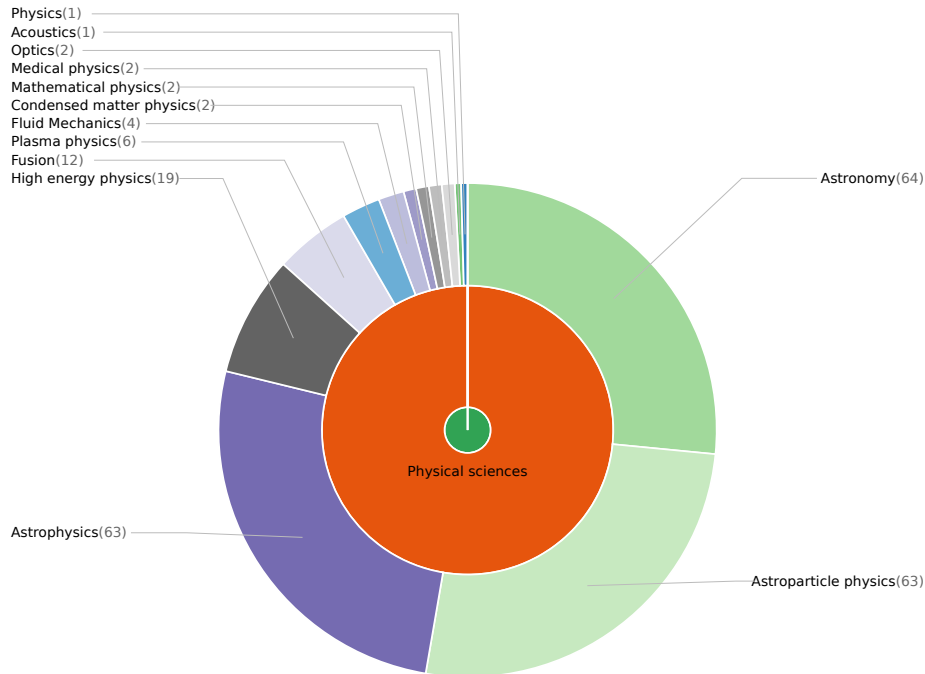}
			\caption{\label{PhysicalSciencesEGIAppDBI}Graph showing the number and distribution of applications in the Physical Sciences domain registered in the EGI Application Database.}
		\end{center}
	\end{figure}

	Often, the porting and integration of new applications is done by a site administrator who attempts to compile and install the researcher's application according to local procedure and policy. This usually leads to a significant amount of lead-time and depends critically on the ability of the site administrator to interpret the needs of the researcher. Furthermore, there can be much ambiguity regarding the resolution of dependencies, optimisation and versions. Here, we come up against a tension between domain expertise and administrative privileges : it is usually the {\it researcher}, not the site administrator who best understands how to compile the applicaiton. However, it is the site administrator who has the knowledge of the local setup which would integrate the applicaiton appropriately and efficiently. It should also be considered that this overhead is multiplied by the number of distinct application stacks which are in use in scientific domains, and exacerbated by the multitude of dependency and optimisation configurations. For the physical sciences domain, this can be appreciated by considering the wide array of applications registered in the EGI Applications Database\cite{EGIAppDB} in Figure \ref{PhysicalSciencesEGIAppDBI}\footnote{For physics alone, there is a complexity and scale which is prohibitive to any particular site administrator. However if one considers the applications maintained in the Homebrew or EasyBuild repositories - $\tilde 600$ and $\tilde 1000$ respectively - the situation seems untenebale.}

	\subsection{CODE-RADE Design Hypotheses}

	CODE-RADE provides a means to transparently integrate a new application into an arbitrary site configuration, in a way which is both user-driven and satisfies the constraints imposed by site administrators. In order to implement the solution, a reconsideration was made of the way in which applications are brought to bear in scientific research and computational infrastructure. This was done as far as possible from first principles, by constructing seven hypotheses of bringing applications to distributed infrastructure. In what follows, we discuss the considerations  of these design hypotheses and how they have influenced the development and results of the project.

	\begin{description}
		\item[Hypothesis 1: Every researcher is an application] \hfill \\
		The core aspect of initial communication between a research infrastructure provider and a researcher is the {\it application}. Every effort should be made then to resolve and understanding of the needs of the researcher down to this level and engage with them on technical aspects of the application.
		\item[Hypothesis 2: No software is an island] \hfill \\
		Every scientific application has a dependency list. No definition of an application can be complete without an explicit expression of the list of its dependencies, which should extend as far as possible down userspace.
		\item[Hypothesis 3: All applications need an environment] \hfill \\
		Just as bioligical organisms require an environment in which to express their functionality, digital organisms (scientific applications) are sterile if they are not placed in an environment capable of expressing them. This environment can be expressed by combinations of compilers, hardware architecture, application dependencies, middleware stacks, {\it etc}.
		\item[Hypothesis 4: There is more than one environment] \hfill \\
		By restricting ourselves to a single environment, we lose main advantages of distributed systems, and erect unncessary barriers to entry for researchers. We should thus design a system which {\it simulates} all of the available sites which applications could in principle be executed, to increase application portability.
		\item[Hypothesis 5: Solutions Decay] \hfill \\
		A ``solution" in this case refers to a software expression of the method for creating an executable version of the application. The exploration of these solutions, including the attempts which failed, are important components of the system knowledge. However, due to the development of software, it is likely that a found solution will not apply for future iterations. Thus, solutions need to be expressed in a machine-readable form {\it which is maintained in a version-controlled repository}
		\item[Hypothesis 6: Humans need not apply\footnote{With apologies to CGP Grey - https://www.youtube.com/watch?v=7Pq-S557XQU}] \hfill \\
		If we assume that the previous hypotheses are true, it stands to reason that most of the work to actuall compile and integrate applications can be automated. This represents a major relief of workload from site administrators, as well as ensuring that human errors are less frequently introduced. Furthermore, if a system is built with respecting open standards, it becomes easier to extend it.
		\item[Hypothesis 7: This is not hard] \hfill \\
		While it may seem that we are proposing a radical shift from common practice, which may seem discouraging, the fact is that continuous integration\cite{ContinuousIntegration} and continuous delivery\cite{ContinuousDelivery} has been common practice in the software development world for several years.
	\end{description}

	Along with these hypotheses, CODE-RADE attempts  to solve problems which impede fluid collaboration and exploitation of computational resources. First of all, the system aims to reduce or remove entire barriers to entry represented by user applications, through the use of continuous integration and delivery, automated builds and version-controlled source code repositories. This allows all involved to know exactly where the application is in its lifecycle, and indeed which build of the application is currently being used.

	Automated continuous delivery ensures that once applications are integrated, they are immediately available at sites which wish to host them, further reducing the workload on site administrators.

	\section{CODE-RADE in detail}

	\subsection{Actors}

	The CODE-RADE platform identifies four kinds of actors which have distinct roles to play :
	\begin{itemize}
		\item Researchers
		\item Research Software Engineers (RSE)
		\item Infrastructure Operators (Ops)
		\item Automated Agents (Bots)
	\end{itemize}

	Researchers are the final users of the application, however are not necessarily the authors of the software. Researchers may not know the details necessary to compile and optimise applications for specific environments, however they are adept in interpreting the results of these applications and are the end consumers of them. The research software engineer, which in some cases is the researcher themselves, thus provides a means to express the application on the various targets in an executable state. Proof positive to the resource providers that applicatons can indeed execute properly on target infrastructures is needed before they will agree to actually host and excute the applications on their sites. Indeed, the characteristics of these target infrastructures are defined by the infrastructure operations team, which know them best. The role of the automtaed agents is to execute the code provided by the research software engineers and see whether it passes the tests set by the infrastructure operations team, at every commit pushed to the repository. This provides a direct link between change in the software and expression of that software (source code version linked to executable version), and since the tests criteria are predefined, once they pass applications can be integrated without human intervention.

	In principle, the entire workflow can be driven by researchers themselves.

	\subsection{Platform}

	In this section, we describe how we have built the CODE-RADE platform, based on the philosophy, hypotheses, and actors described above.

	\subsubsection{Components}

	The implementation of CODE-RADE consists essentially of three generic components :

	\begin{itemize}
		\item Continuous integration tool
		\item Version controlled software repository
		\item Automated software delivery tool
	\end{itemize}

	There are several choices  which one could use to build a similar platform to CODE-RADE by choosing a particular tool to fulfill each component. In our case, we have selected Jenkins\cite{Jenkins}, GitHub\cite{Github} and CERN Virtual Machine Filesystem\cite{CVMFS}. The reasoning and justification for these specific choices is given in some more detail below in section \ref{Discussion}. The components have specific roles to play in each phase of the CODE-RADE workflow, which we describe next.

	\subsubsection{CODE-RADE workflow}

	Although the main aim of CODE-RADE is to ensure that applicaitons are available for execution on arbitrary remote sites, there are collateral issues which have importance to a community. We identify several goals in CODE-RADE which have their own flow of actions and we thus define the generic workflow for application {\bf integration, delivery, review} and {\bf publication}.

	The most common workflow is simply stated as :

	\begin{enumerate}[label=\arabic*]
		\item Build
		\item Test
		\item Deliver
	\end{enumerate}

	The first step (build) this refers to the synchronisation of the source code with relevant updates. This is initiated by the Research Software Engineer: a commit is made to a repository, detailing a change which the RSE authors. This triggers an event on Jenkins, which pulls the latest commit and executes the scripts in the repository which will build and test the code. The second step (Test) is conditionally executed on success of the previous step. This step checks the viability of the built application and executes whatever tests are specified, then installs the application into the integration environment. Finally the third step (Deliver) cleans the built application and installs it into the deploy environment, then ships the executable code to the repository for subsequent replication.

	Although there are several tools for building widely-used applications (see Section \ref{Discussion} below), we have chosen a convention within the project to have three scripts which perform the compilation, functional testing and integration into the delivery environment respectively. These have a pattern of phases :

  \begin{description}
    \item[Build - \texttt{build.sh}] does the following :
      \begin{enumerate}[label=\arabic*]
        \item Ensure that the source bundle for the requested application is available and up to date locally.
        \item Configure the build (usually using traditional AutoTools \cite{AutoTools} or CMake \cite{CMake} : \texttt{./configure \&\& make}). This phase is intended to determine whether the application has the relevant dependencies already included in the repository, and whether it can be compiled with the availabe compilers and dependencies.
      \end{enumerate}
    \item[Test - \texttt{check-build}] does the following :
      \begin{enumerate}[label=\arabic*]
        \item Execute any internal functional tests that the application may contain.
        \item Execute integration tests that the Ops team specifies
        \item Install the application into the integration environment - this ensures that the compiled application (binaries and libraries) are available for dependent applications, and a modulefile is created to easily add the application to a shell.
        \item Execute (in the integration environment) any researcher-specified tests, referred to a minimum viable execution.
      \end{enumerate}
    \item[Deliver - \texttt{deploy.sh}] does the following :
      \begin{enumerate}[label=\arabic* ]
        \item Clean the build and prepare for recompilation using the deploy environment
        \item Recompile using the deploy environment
        \item Install into the deploy environment
        \item Create the module file for the application.
      \end{enumerate}
  \end{description}


Each job as defined on Jenkins consists of at least these three tasks. If each passes successfully, a job to update the CVMFS repository is triggered, which puts the repository into transaction, pulls in the newly built applications and module files, then publishes a new version of the repository. This is automatically and transparently available at sites after a few seconds, once the CVMFS client detects the change in the repository hash. In order to make it easier for end users or client applications to check which version of the repository they are using\footnote{This can be determined more accurately using the cvmfs client tools, but this may not be intuitive to users.}, an integer increment is added to a file in the repository which is linked to the job which resulted in the latest version.

In short, a commit to a repository by a researcher or RSE will result in the application being automatically delivered to arbitrary endpoints, as long as quality and functional tests set by Ops and researchers respectively are passed.

	\section{Discussion}\label{Discussion}

Our particular combination of existing tools (Github, Jenkins and CVMFS) has resulted in a powerful platform with which to exploit any existing computational infrastructure, in an automated user-driven way. Some aspects which together make CODE-RADE a novel and powerful platform should be highlighted.

\subsection{CODE-RADE is cross-platform}

CODE-RADE takes a software description of an scientific tool and builds a digital expression which can be executed on arbitrary architectures. This makes CODE-RADE different from more traditional package managers, since these only build for the architecture on which the user is currently. By simulating the execution environment which may in principle be available at geographically distributed sites. We currently define a target according to a matrix of characteristics, which are encoded as variables in the build and deploy environments :

\begin{description}
	\item[ARCH]: The CPU architecture of the execution environment
	\item[OS]: The operating system of the execution environment
	\item[SITE]: A meta-variable encompassing aspects of specialised hardware (GPU, Myrinet, GPFS, etc) and other aspects (software licenses, available middleware etc.)
\end{description}

For each combination of these axes, specific optimisations can be applied, ensuring that the final result (compiled application) is both properly compiled, and will execute efficiently on the target site
By building on as many environments as possible, for as many combinations as possible, CODE-RADE promotes diversity in computing platforms, without exacerbating the paradox of plentiful computing.

\subsection{CODE-RADE is atomic}

Fine-grained control over dependencies, versions and targets
Relevant action taken on each event

\subsection{CODE-RADE is community-based}

No restriction on the applications that can be integrated.
Anyone can contribute applications, resources, code review, etc

\subsection{CODE-RADE is Automated}

Heavy reliance on automated agents to reduce bias, lead time
User-driven

	\subsection{CODE-RADE for Physics}

	ok, physicist - your time to shine. Here we say how CODE-RADE makes running physics applications so much easier on a distributed platform.

Problem area : nuclear physics
Simulation of nuclear phenomena
Ithemba LABS/GSI
Integration time – 21 hours (no prior experience)

	\section{Summary, Conclusion and Futher work}

Making the best use of computational infrastructure comes down to running applications
Maintaining and porting them is hard and tricky. Communication is the hardest part.
We've built an automated porting system, which will deliver functional, tested, relevant software to your site. Can ease the communication blocks in collaborative work
Come on in and help us build it.

\section*{Acknowledgements}
The authors would like to thank contributors to the project for ideas, discussion and code
The original idea co-developed by one of the authors (BB) and Fanie Riekert (University of the Free State). Input and critique was provided  by Dane Kennedy and Sakhile Masoka (Centre for High Performance Computing, Cape Town) and Peter van Heusden (South African National Bioinformatics Institute). The authors recognise the technical support of the EGI CVMFS Task Force support, in particular Catalin Condurache (European Grid Initiative and Science and Technology Facilities Council, UK). Several discussions on CODE-RADE design and extension were had with Timothy Carr (University of Cape Town e-Research Centre).

CODE-RADE is supported by the Sci-GaIA project under grant 654237 of the European Commission's Horizon 2020 programme

\end{document}